\documentclass{iau} 
\usepackage{graphicx}
\usepackage{natbib}
\usepackage{journals}
\usepackage[colorlinks=true,linkcolor=blue,anchorcolor=blue,citecolor=blue,filecolor=blue,pagecolor=blue,urlcolor=blue]{hyperref}

\title[Targeted millisecond pulsar surveys of \textit{Fermi} $\gamma$-ray sources with LOFAR]
{Targeted millisecond pulsar surveys of \textit{Fermi} $\gamma$-ray sources with LOFAR}

\author[C.~G.~Bassa et al.]{C.~G.~Bassa$^1$, Z.~Pleunis$^2$,
  J.~W.~T.~Hessels$^{1,3}$, E.~C.~Ferrara$^{4,5}$,
  V.~I.~Kondratiev$^{1,6}$, S.~Sanidas$^3$, A.~G.~Lyne$^7$,
  B.~W.~Stappers$^7$, S.~M.~Ransom$^8$ and the \textit{Fermi} Pulsar
  Search Consortium }

\affiliation{$^1$ASTRON, the Netherlands Institute for Radio
  Astronomy, Postbus 2, NL-7990 AA Dwingeloo, The Netherlands;
  \texttt{bassa@astron.nl}\\[\affilskip]$^2$Department of Physics and
  McGill Space Institute, McGill University, 3600 University St.,
  Montreal, QC H3A 2T8, Canada\\[\affilskip]$^3$Anton Pannekoek
  Institute for Astronomy, University of Amsterdam, Science Park 904,
  1098 XH Amsterdam, The Netherlands\\[\affilskip]$^4$Center for
  Research and Exploration in Space Science, NASA Goddard Space Flight
  Center, Greenbelt, MD 20771, USA\\[\affilskip]$^5$Department of
  Astronomy, University of Maryland, College Park, MD 20742,
  USA\\[\affilskip]$^6$Astro Space Centre, Lebedev Physical Institute,
  Russian Academy of Sciences, Profsoyuznaya Str.\ 84/32, Moscow
  117997, Russia\\[\affilskip]$^7$Jodrell Bank Centre for
  Astrophysics, The University of Manchester, Manchester, M13\,9PL,
  United Kingdom\\[\affilskip]$^8$National Radio Astronomy
  Observatory, Charlottesville, VA 22903, USA}

\pubyear{2017}
\volume{337}
\jname{Pulsar Astrophysics -­ The Next 50 Years}
\editors{P. Weltevrede, B.B.P. Perera, L. Levin Preston \& S. Sanidas, eds.}

\def\arcmin{\hbox{$^{\prime}$}}

\def\lta{\mathrel{\hbox{\rlap{\hbox{\lower4pt\hbox{$\sim$}}}\hbox{$<$}}}}
\def\gta{\mathrel{\hbox{\rlap{\hbox{\lower4pt\hbox{$\sim$}}}\hbox{$>$}}}}

\def\degr{\hbox{$^\circ$}}
\usepackage{upgreek}

\begin{document}

\maketitle

\begin{abstract}
  We have used LOFAR to perform targeted millisecond pulsar surveys of
  \textit{Fermi} $\gamma$-ray sources. Operating at a center frequency
  of 135\,MHz, the surveys use a novel semi-coherent dedispersion
  approach where coherently dedispersed trials at coarsely separated
  dispersion measures are incoherently dedispersed at finer
  steps. Three millisecond pulsars have been discovered as part of
  these surveys. We describe the LOFAR surveys and the properties of
  the newly discovered pulsars.  
\end{abstract}

\firstsection 
\section{Introduction}
The Large Area Telescope (LAT) on the \textit{Fermi} $\gamma$-ray
Space Telescope has given us an unprecedented view of the $\gamma$-ray
sky. It has higher sensitivity and spatial resolution than previous
$\gamma$-ray observatories, and as a result, over three thousand
$\gamma$-ray point sources have been detected in the first 4\,years of
data \citep{aaa+15_3fgl}. The majority of these sources are associated
with different types of $\gamma$-ray emitting active galaxies, while a
significant fraction (about 9\%) are identified as pulsars,
particularly energetic millisecond pulsars (MSPs). Radio pulsar
surveys of unidentified \textit{Fermi} $\gamma$-ray sources have been
a very fruitful way for discovering new MSPs (see \citealt{rap+12} and
references therein), and these surveys have been crucial in
determining the nature of the presently unidentified population of
$\gamma$-ray sources.

To date, targeted MSP surveys of unidentified \textit{Fermi}
$\gamma$-ray sources have performed at observing frequencies above
300\,MHz (e.g.\ \citealt{cgj+11,hrm+11,kjr+11,rrc+11}). Since the
radio emission from pulsars generally exhibit steep spectra
\citep{mkkw00,blv13}, these survey may potentially miss MSPs with very
steep spectra. Here, we present the results of targeted surveys for
MSPs towards \textit{Fermi} $\gamma$-ray sources with LOFAR, using a
semi-coherent dedispersion approach to limit the effects of
dispersion.

\section{Survey description}
We used LOFAR to perform two targeted surveys for radio MSPs towards
\textit{Fermi} $\gamma$-ray sources. The first, a pilot survey,
targeted 52 unassociated $\gamma$-ray sources from the 3FGL catalog by
\citet{aaa+15_3fgl}, while the second survey used unpublished data to
select 72 \textit{Fermi} $\gamma$-ray sources with pulsar-like
spectral and variability properties. All sources were located outside
of the Galactic plane with $|b|>10\degr$, visible to LOFAR (elevations
$>30\degr$) and well localized (95\% confidence uncertainty region
less than $10\arcmin$ in diameter). All sources were observed with the
high-band antennas (HBAs) of 21 LOFAR core stations \citep{hwg+13},
where the LOFAR beamformer was configured to form 7 tied-array beams,
covering a $10\arcmin$ diameter field-of-view, each producing
dual-polarization, complex, Nyquist sampled timeseries over 39\,MHz of
bandwidth centered at a frequency of 135\,MHz. Integration times of
20\,min were used in all cases, using single observations for the
pilot survey, and two observations, separated by a few days, for the
second survey. The two separated observations were aimed to improve
the probability of detecting short period or eclipsing binary pulsar
systems.

To limit smearing due to dispersion, we used a semi-coherent
dedispersion approach, where the complex voltage data was first
coherently dedispersed to 80 trial dispersion measures
($\mathrm{DM}$s) at steps of 1\,pc\,cm$^{-3}$ from 0.5 to
79.5\,pc\,cm$^{-3}$ using the GPU accelerated \texttt{cdmt} software
\citep{bph17}. For the pilot survey the resulting filterbanks had time
and frequency resolution of 40.96\,$\upmu$s and 24.41\,kHz,
respectively, while the second survey reduced the resolution to
81.92\,$\upmu$s and 48.82\,kHz to improve processing times, without
significantly reducing sensitivity. A GPU accelerated brute force
incoherent dedispersion algorithm \citep{bbbf12} was used to
dedisperse the filterbank files at steps of 0.002\,pc\,cm$^{-3}$ from
$-0.5$ to $0.5$\,pc\,cm$^{-3}$ around the coherent $\mathrm{DM}$. All
of the 40\,000 dedispersed timeseries were searched for periodic
signals using frequency domain acceleration searching, using a GPU
accelerated algorithm implemented in \texttt{PRESTO}
\citep{ran01,rem02}. Standard sifting and folding tools from the
\texttt{PRESTO} suite were used to assess pulsar candidates.

The survey sensitivity is estimated at about 2\,mJy for millisecond
spin periods \citep{pbh+17}. The semi-coherent dispersion method
corrects for the smearing due to dispersion, and hence the survey
sensitivity is ultimately limited by multi-path scattering. The
\citet{bcc+04} scattering relation predicts scattering to become
dominant above $\mathrm{DM}$s of 40 and 80\,pc\,cm$^{-3}$ for 1 and
10\,ms spin periods, respectively \citep{pbh+17}. Since these surveys
specifically target sources at high Galactic latitude, on average, the
maximum $\mathrm{DM}$ is less than 50\,pc\,cm$^{-3}$, and smearing due
to scattering is not expected to be a dominant effect.

\begin{figure}[!t]
  \begin{center}
    \includegraphics[width=\textwidth]{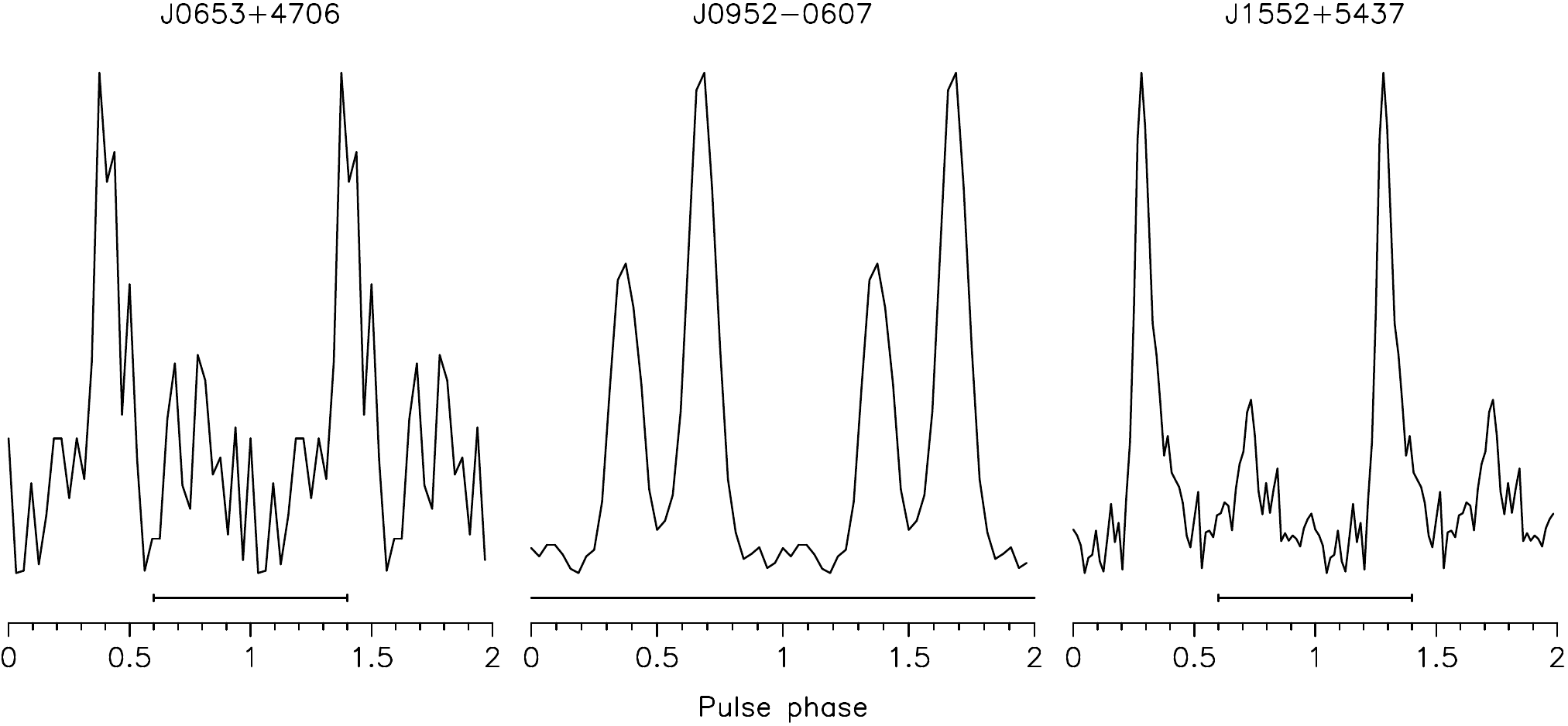} 
    \caption{Pulse profiles taken from the discovery observations. Two
      pulse periods are shown. The horizontal error bar denotes the
      dispersive smearing in the absence of coherent dedispersion. }
    \label{fig:profiles}
  \end{center}
\end{figure}

\section{Pulsar discoveries}
The LOFAR surveys described here resulted in the discovery of three
new radio MSPs. Their pulse profiles are shown in
Figure\,\ref{fig:profiles}. The first, PSR\,J1552+5437, was found in
the pilot survey observation of $\gamma$-ray source
3FGL\,J1553.1+5437, and is an isolated pulsar with a 2.43\,ms spin
period at a $\mathrm{DM}$ of 22.90\,pc\,cm$^{-3}$. Folding of the
7\,yr of \textit{Fermi} LAT $\gamma$-ray photons yields a phase
coherent timing ephemeris, and a $\gamma$-ray pulse profile that
appears to be aligned with that in the radio. Non-detections of the
pulsar at 820\,MHz and 1.5\,GHz indicate PSR\,J1552+5437 has a steep
radio spectrum, for $S_\nu\propto\nu^\alpha$ with
$\alpha<-2.8(0.4)$. This discovery has been published by
\citet{pbh+17}.

PSR\,J0952$-$0607 was the second LOFAR MSP discovery. With a spin
period of 1.41\,ms and $\mathrm{DM}=22.41$\,pc\,cm$^{-3}$, it is the
second fastest spinning pulsar known. Its spin frequency of
$\nu=707$\,Hz is 9\,Hz slower than the fastest MSP discovered,
Terzan\,5ad ($\nu=716$\,Hz; \citealt{hrs+06}). The pulsar is in a
6.42\,hr binary system with an optically detected \textit{black widow}
companion. Remarkably, even at the low observing frequencies of LOFAR,
no radio eclipses are seen. Like PSR\,J1552+5437, J0952$-$0607 also
has a steep radio spectrum, with $\alpha\sim-3$, being detectable at
150\,MHz with LOFAR and 350\,MHz with the GBT. This discovery has been
published by \citet{bph+17}. Analysis of the \textit{Fermi}
$\gamma$-ray data is ongoing.

The third MSP discovered by LOFAR is PSR\,J0653+4706. It has a
4.75\,ms spin period and a $\mathrm{DM}$ of 25.54\,pc\,cm$^{-3}$. The
pulsar is a member of a 5.84\,d binary system with a
$M_\mathrm{c}\gta0.21$\,M$_\odot$ companion. PSR\,J0653+4706 is also
visible with the GBT at 820\,MHz and at 1.5\,GHz with the Lovell
telescope (under favorable scintillation), indicating the radio
spectrum of PSR\,J0653+4706 is not as steep as the other two LOFAR
discoveries. Its location on the sky may make it a suitable addition
to pulsar timing arrays.

\section{Discussion}
The discovery of these three radio MSPs, being the first discovered
with digital aperture arrays through their pulsations, shows the
promise of pulsar surveys at low frequencies with LOFAR, the MWA, LWA
and SKA1-Low. The crucial ingredient in their discovery is the use of
coherent dedispersion to remove dispersive smearing.

The steep radio spectra of PSRs\,J0952$-$0607 and J1552+5437 add to
the emergent picture in which the fastest spinning pulsars have the
steepest spectra. Recent studies with LOFAR \citep{kvh+16} and the
GMRT \citep{fjmi16} show that the majority of MSPs with spin
frequencies larger than $\nu>300$\,Hz ($P<3.33$\,ms) have spectra
steeper than $\alpha<-2.5$. Furthermore, \citet{kvl+15} and
\citet{fjmi16} found that MSPs seen in $\gamma$-rays tend to have
steep radio spectra. These systems typically have aligned radio and
$\gamma$-ray profiles \citep{egc+13,jvh+14}, as seems the case for
PSR\,J1552+5437 \citep{pbh+17}. It is suggestive that these tendencies
are pointing to a commonality, possibly related to the small light
cylinder of fast spinning MSPs, forcing the co-location of
$\gamma$-ray and steep spectrum radio emission.

After presenting these results at the IAU337 symposium there was some
discussion on whether the tendency that the fastest spinning MSPs have
the steepest radio spectra is an observational bias in their discovery
or an intrinsic property of these pulsars. As an argument in favor of
a discovery bias, Terzan~5ad, the fastest spinning MSP ($\nu=716$),
was mentioned as it was discovered and timed at S-band
\citep{hrs+06,prf+17}. Arguments against the steep spectra being a
bias in discovery revolved around the inherent difficulty, due to
scattering and dispersion, of discovering fast spinning MSPs at low
versus high observing frequencies. As part of this discussion, Scott
Ransom was willing to bet that the tendency is due to observational
bias. Scott's bet was taken on by Joeri van Leeuwen, where the loser
shall make the winner a bottle of steeped liquor (such as a
citrus-infused gin or bacon-steeped vodka), if the majority of the
tendency is confirmed as being an intrisic MSP property or an
observational within 5\,years of the IAU337 symposium.

\bibliographystyle{aasjournal}

\end{document}